\documentclass[
]{ceurart}

\usepackage{listings}
\usepackage{enumitem}
\begin{document}

\copyrightyear{2023}
\copyrightclause{Copyright for this paper by its authors.
  Use permitted under Creative Commons License Attribution 4.0
  International (CC BY 4.0).}

\conference{ER2023: Companion Proceedings of the 42nd International Conference on Conceptual Modeling: ER Forum, 7th SCME, Project Exhibitions, Posters and Demos, and Doctoral Consortium, November 06-09, 2023, Lisbon, Portugal}

\title{Conceptual model interpreter for Large Language Models}


\author[1]{Felix Härer}[%
orcid=0000-0002-2768-2342,
email=felix.haerer@unifr.ch,
url=https://www.unifr.ch/inf/digits/en/group/team/haerer.html,
]
\address[1]{University of Fribourg,
  Boulevard de Pérolles 90, 1700 Fribourg, Switzerland}



\begin{abstract}
  Large Language Models (LLMs) recently demonstrated capabilities for generating source code in common programming languages. Additionally, commercial products such as ChatGPT 4 started to provide code interpreters, allowing for the automatic execution of generated code fragments, instant feedback, and the possibility to develop and refine in a conversational fashion. With an exploratory research approach, this paper applies code generation and interpretation to conceptual models. The concept and prototype of a conceptual model interpreter is explored, capable of rendering visual models generated in textual syntax by state-of-the-art LLMs such as Llama~2 and ChatGPT 4. In particular, these LLMs can generate textual syntax for the PlantUML and Graphviz modeling software that is automatically rendered within a conversational user interface. The first result is an architecture describing the components necessary to interact with interpreters and LLMs through APIs or locally, providing support for many commercial and open source LLMs and interpreters. Secondly, experimental results for models generated with ChatGPT 4 and Llama 2 are discussed in two cases covering UML and, on an instance level, graphs created from custom data. The results indicate the possibility of modeling iteratively in a conversational fashion.
\end{abstract}

\begin{keywords}
  Large Language Model \sep
  Conceptual Model \sep
  Code Generation \sep
  Interpreter
\end{keywords}

\maketitle

\section{Introduction}
\label{introduction}

In recent years, Large Language Models (LLMs) have seen broad adoption due to the wide variety of successful applications utilizing transformers~\cite{VaswaniAttention2017}, enabling language-related tasks at higher abstraction levels, e.g., when detecting and describing images, performing audio tasks such as speech-to-text and voice cloning, or handling text for modeling topics, summarization, or translation~\cite{XuLearning2023}. Especially in the form of Generative Pre-trained Transformer (GPT) models, language analysis and generation capabilities evolved and today include programming languages~\cite{RayChatGPT2023,MaScope2023} in addition to first indications of support for domain-specific languages and conceptual modeling~\cite{FillChatGPT2023}. After the large-scale adoption of ChatGPT 4, gaining over 100 million active users within two months after its initial release in November 2022~\cite{ChowWhy2023}, commercially available products such as ChatGPT 4 and the GitHub Copilot started to introduce further features for software development in 2023~\cite{Wermelinger2023}. Notably, the conversational generation of source code is now complemented by the ability to invoke external APIs using Plugins~\cite{OpenAIPlugins2023} and the execution of generated source code with a code interpreter~\cite{DavisTesting2023}. In particular, OpenAI introduced a code interpreter that allows a user in a conversation to describe or specify a task for programming, automatically generating the source code, and automatically executing it instantly to provide output and computation results as a response. Examples demonstrated capabilities to interpret code fragments, e.g., calculating functions or creating plots in Python. The conversational approach also provides the potential to include code generation with instant feedback and step-wise refinement into the software development process, allowing for the iterative development of code fragments with instant feedback on execution results.

In this paper, the conversational approach with instant feedback and step-wise refinement is applied to conceptual models. For exploring the potential of this approach, the research objectives are (1.) to determine whether the approach can be realized with state-of-the-art commercial or open source LLMs and interpreters, and (2.) how a possible realization could be constructed in terms of an architecture. By exploratory research, these objectives are investigated through the construction of a prototype chat application. 

Potentially, conceptual models may be created within a dialogue where a modeling task is specified fully or partially in natural language, textual syntax for models is then generated by the LLM and automatically rendered visually by interpreters, e.g., for PlantUML or Graphviz syntax. In addition, the interpreter is a first step towards specifying executable models for generating software through LLMs that can be executed directly in local environments or at the server side, e.g., generating step function models for cloud platforms and blockchains~\cite{Haerer2023}. 

\begin{figure*}[ht]
    \centering
    \includegraphics[width=0.82\linewidth]{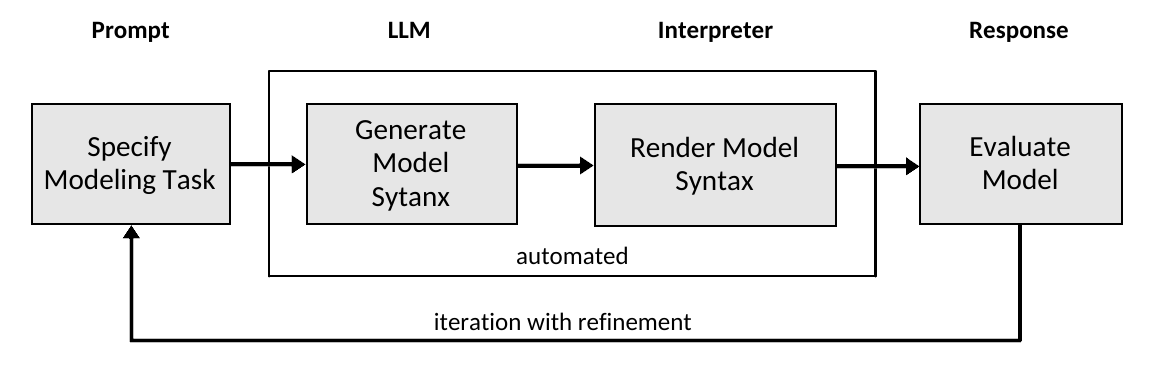}
    \caption{Iterative modeling process from a user perspective, generating and rendering the concrete syntax of a model in a conversational fashion.
    }
    \label{fig:process}
    \vspace{-2mm}
\end{figure*}

Modeling with instant feedback allows for an iterative process with refinement as described in Figure~\ref{fig:process}. Such a process could be applied generally in many areas, e.g., in requirements and software engineering, where the generation of structure or behavior diagrams is above the level of source code. Further, it lowers the barrier to entry for modeling and design activities. While LLMs also lower the barrier to generating source code directly, the difficulty oftentimes lies in its evaluation; that is, to evaluate whether source code and the implied software design are suitable and fulfill the requirements. 

The remainder of this paper is structured as follows. Section~\ref{background-related-work} introduces background on LLMs and related work on applying LLMs for conceptual modeling. In Section~\ref{cm-interpreter}, the concept of a model interpreter and a possible realization architecture are outlined. Section~\ref{results} discusses experimental results of the prototype with prompts and generated models for ChatGPT~4 and Llama~2. Section~\ref{conclusion} outlines the overall results and concludes.

\section{Background and related work}
\label{background-related-work}

\subsection{Background}

Generative Pre-trained Transformer models (GPT) are a class of Large Language Models (LLMs) that apply a specific architecture on the principle of predicting or completing text, progressing on a sequence of tokens.

An input sequence is given by a prompt and the context of a dialogue preceding it, where words are split into tokens, encoded, and represented by a vector according to an embedding in a high-dimensional vector space in the thousands of dimensions~\cite{RayChatGPT2023,RothmanTransformer2022,VaswaniAttention2017}. With an additional encoding of the position in the sequence, each token is applied to a transformer architecture to generate candidates for the next token with a probability distribution. Tokens pass sequentially through multiple transformer layers of transformer blocks, each consisting of attention and feed-forward network components. Each transformer block produces tokens and probabilities with increasing abstraction. Within a block, attention directs the focus to select tokens seen before within a similar context, based on similar attention at prior positions in the sequence. Attention mechanisms achieve this by calculating attention at a given position relative to other positions in the sequence~\cite{VaswaniAttention2017}, instead of relying only on co-occurrences. The output of a transformer block with tokens and a probability distribution is dependent on the weights of the feed-forward network and normalization. After passing all the layered transformer blocks, 96 in the case of GPT-3~\cite{IsaevScaling2023}, the next token is ultimately selected according to probability with a sampling algorithm~\cite{Labonne2023} and appended to the sequence. Subsequently, the iterative token generation produces the response and becomes part of the context. 

LLMs notably differ in terms of their supported context size, allowing for 4096 tokens in Llama~2~\cite{TouvronLlama2023}, 8192 in the standard version of GPT-4, and 32768 in its extended version\footnote{\url{https://platform.openai.com/docs/models}}. Substantial differences also exist in the number of parameters applied by the training data with 7, 13, 34, or 70 billion parameters for Llama~2 variants~\cite{TouvronLlama2023}, 175 billion in GPT-3.5~\cite{RayChatGPT2023}, and possibly trillions of parameters in GPT-4~\cite{RayChatGPT2023}. Exact specifications, also concerning the transformer architecture, are unknown for the closed-source GPT-4 model. Further optimizations are applied for LLMs used in a conversational fashion, e.g. GPT-4 and Llama~2, to adapt responses to the style of conversational dialogues and to emphasize or suppress specific context by applying Reinforcement Learning from Human Feedback (RLHF) and fine-tuning~\cite{RayChatGPT2023}. 


\subsection{Related work}

While prior work on LLMs and modeling applications is scarce, first publications do exist for conceptual modeling, business process management, and software modeling, in addition to related findings for programming languages.

For conceptual modeling, the application of ChatGPT has been investigated before in a publication by Fill et al.~\cite{FillChatGPT2023} which suggested the generation of UML class diagrams in PlantUML syntax as well as creating ER, workflow, and Heraklit models in a custom syntax. Results further demonstrate the capability of GPT-4 to generate new models beyond examples potentially existing in training data, by issuing prompts for textbook-like case descriptions and abstract examples. This paper is inspired by this investigation and seeks to integrate multiple LLMs and interpreters in a more generalized way to explore their modeling capabilities towards further systematic evaluations in the future.

For Business Process Management (BPM), Vidgof et al.~\cite{Vidgof2023} note opportunities and challenges along the BPM lifecycle. In relation to modeling, discovery is a major topic for discovering process models through process mining. While LLMs can support discovery generally, e.g. from documentation, the generation of BPMN in XML format is noted as well as parsing existing XML process models with a LLM, opening the potential to query the LLM for knowledge on the syntax, the semantics of the process, and potential execution behavior. Regarding process implementation, BPMN models might be augmented with plain text, and accessed in a user-specific way through chatbots. 

For software modeling, Cámara et al.~\cite{Camara2023} provide an experience report on using ChatGPT for UML modeling tasks in different notations. ChatGPT produced diagrammatic notation using characters, was found to support the PlantUML and USE notations ``generally well", and could also generate OCL expressions. For the test cases, ChatGPT seemed to handle the creation of classes with attributes in addition to associations, aggregations and compositions, simple inheritance and role names of association ends. Certain elements required explicit indication, e.g., enumerations, and results using abstracts and association classes were not acceptable. Further findings indicate results were generally correct, could exhibit small syntactic errors, had high variability for test cases and randomness over time, and depended on the domain understanding of ChatGPT. The paper also notes size limitations of about 8 to 10 classes given a single prompt and the possibility to construct larger models iteratively in further prompts.

Further limitations regarding syntax and semantics aspects were investigated for programming languages in a systematic study by Ma et al.~\cite{MaScope2023}. It finds ChatGPT generally ``excels at understanding code syntax (AST)" while struggling with semantics. In particular, it finds ChatGPT generally understands syntax, is capable of inference based on an abstract syntax tree (AST), and can perform static analysis. However, the understanding of semantics and dynamic behavior was found limited. For static analysis and semantic aspects, ChatGPT also seemed prone to hallucination, i.e., generating non-existent facts, in some cases.

\section{Conceptual model interpreter}
\label{cm-interpreter}

This section outlines the overall concept together with requirements, leading to the discussion of the architecture for describing the system structure.

\subsection{Concept and requirements}

The concept of a model interpreter is first explored from a requirements perspective. This discussion aims at introducing the overall concept and stating requirements that need to be fulfilled in order to construct a corresponding architecture. 

\subsubsection*{Requirements} 


\begin{itemize}[leftmargin=!,labelindent=0pt,itemindent=0pt]

    \item[1.] Conversational user interface. The application requires a conversational user interface, where a continuously ongoing dialogue is presented between the user, a LLM, and an interpreter. A user enters a textual prompt, describing a modeling task, that is appended to the dialogue and sent to a LLM. The generated LLM response in text form, potentially containing a concrete syntax of a modeling language, is appended to the dialogue. In case the response contains the concrete syntax of a known modeling language, it is sent to an interpreter. The execution result of the interpreter is appended to the dialogue in text form and, if a visual model could be rendered, in an image format. 
    
    \item[2.] LLM inference. For running LLM inference, the selection of a suitable LLM, parametrization, and the local or remote execution of the inference are required.
    
    \begin{itemize}[leftmargin=!,labelindent=0pt,itemindent=0pt]
    
        \item[2.1]LLM Selection. For supporting multiple open source and commercial LLMs, possibly differing in their capabilities of generating concrete modeling language syntax, a LLM needs to be selected for inference.

        \item[2.2] LLM Parametrization. Depending on the selected LLM, setting hyperparameters might be required for inference, e.g., to influence the predictability and stability of responses for given prompts. Typically, open source LLMs provide fine-grained control over the sampling in the sequence of tokens, often depending on \emph{temperature}, with lower values decreasing randomness when choosing the next token randomly from the candidates, \emph{top\_k}, limiting the random choice of a token to the $k$ most probable candidates, and \emph{top\_p}, limiting a token to be chosen from the most probable candidates as far as they are, in sum, within probability $p$~\cite{Labonne2023}. 
        
        \item[2.3] Local LLM Inference. LLMs require a runtime component for running inference locally. Especially open source LLMs tend to be suitable for execution within a client application on end-user devices, depending on computation, storage, and memory requirements.
        
        \begin{itemize}[leftmargin=!,labelindent=5.9pt,itemindent=0pt]

            \item[2.3.1] LLM Runtime. A runtime is required for a specific LLM, depending on its architecture and the support of required software or hardware inference or acceleration components. For example, the open source runtime llama.cpp is a C and C++ implementation supporting well-known models such as Alpaca, Vicuna, Falcon, and Llama~2 by inference on x86 and ARM CPUs, e.g. by AVX instructions and the ARM instructions through the Metal framework, as well as on GPUs through CUDA\footnote{\url{https://github.com/ggerganov/llama.cpp}}.
                        
            \item[2.3.2] LLM Files. LLMs might be provided in versions differing in formats, quantization, and overall model size, mostly due to the number of parameters that are typically in the billions. The runtime is required to support and manage LLMs and file formats given sufficient storage and memory capacity. With the released versions in 7, 13, and 70 billion parameters, Llama~2 ranges between 14 GB and 138 GB\footnote{\url{https://huggingface.co/meta-llama}}.

        \end{itemize}
        
        \item[2.4] Remote LLM Inference. LLMs that are available remotely, especially commercial products not available as open source software, require an API client connected to compatible remote servers.
        
        \begin{itemize}[leftmargin=!,labelindent=5.9pt,itemindent=0pt]
        
            \item[2.4.1] LLM API Client. Common APIs include the OpenAI API\footnote{\url{https://platform.openai.com/docs/api-reference/chat}}, used for ChatGPT and many commercial and non-commercial LLMs, and APIs of cloud platforms such as the Replicate HTTP API\footnote{\url{https://replicate.com/docs/reference/http}}. Oftentimes, HTTP and REST are used with specialized bindings for common programming languages.
            
            \item[2.4.2] LLM Server. Outside the client application, LLM servers provided, e.g., by OpenAI, Microsoft Azure, Replicate, or Google are part of the system architecture. Implications, at least to privacy, cost, and performance, need to be considered in this regard.
            
        \end{itemize}
    \end{itemize}

    \item[3.] Interpreter. For interpreting models, the selection, parametrization, and a runtime are required.
    
    \begin{itemize}[leftmargin=!,labelindent=0pt,itemindent=0pt]
        \item[3.1] Interpreter Selection. To support the concrete syntax of multiple modeling languages in a textual format, a suitable interpreter needs to be selected. E.g., an interpreter for PlantUML\footnote{\url{https://plantuml.com/}} such as Plantweb\footnote{\url{https://plantweb.readthedocs.io/}}.
        
        \item[3.2] Interpreter Parametrization. Depending on the interpreter, parameters might be required for determining the rendering layout or the output format. While a textual syntax is assumed for the input, the output might result in vector graphics such as SVG, raster graphics, or textual drawings~\cite{Camara2023}.
        
        \item[3.3] Interpreter Runtime. For rendering visual models of a specific modeling language, an interpreter supporting a concrete syntax in textual format is required. The interpreter needs to be capable of processing the textual input with low latency, rendering a visual model as an output.
        
    \end{itemize}
    
    \item[4.] Data Store. For assessing the results of different LLMs and interpreters over the course of multiple conversations, a data store is required. That is, the selected LLM, interpreter, LLM parameters, and interpreter parameters need to be stored for a specific conversation in addition to the requested prompts, generated responses, and rendered visual models.
    
    \medskip
\end{itemize}

\subsection{Architecture}

\begin{figure*}[ht]
    \centering
    \vspace{-2mm}
    \includegraphics[width=0.98\linewidth]{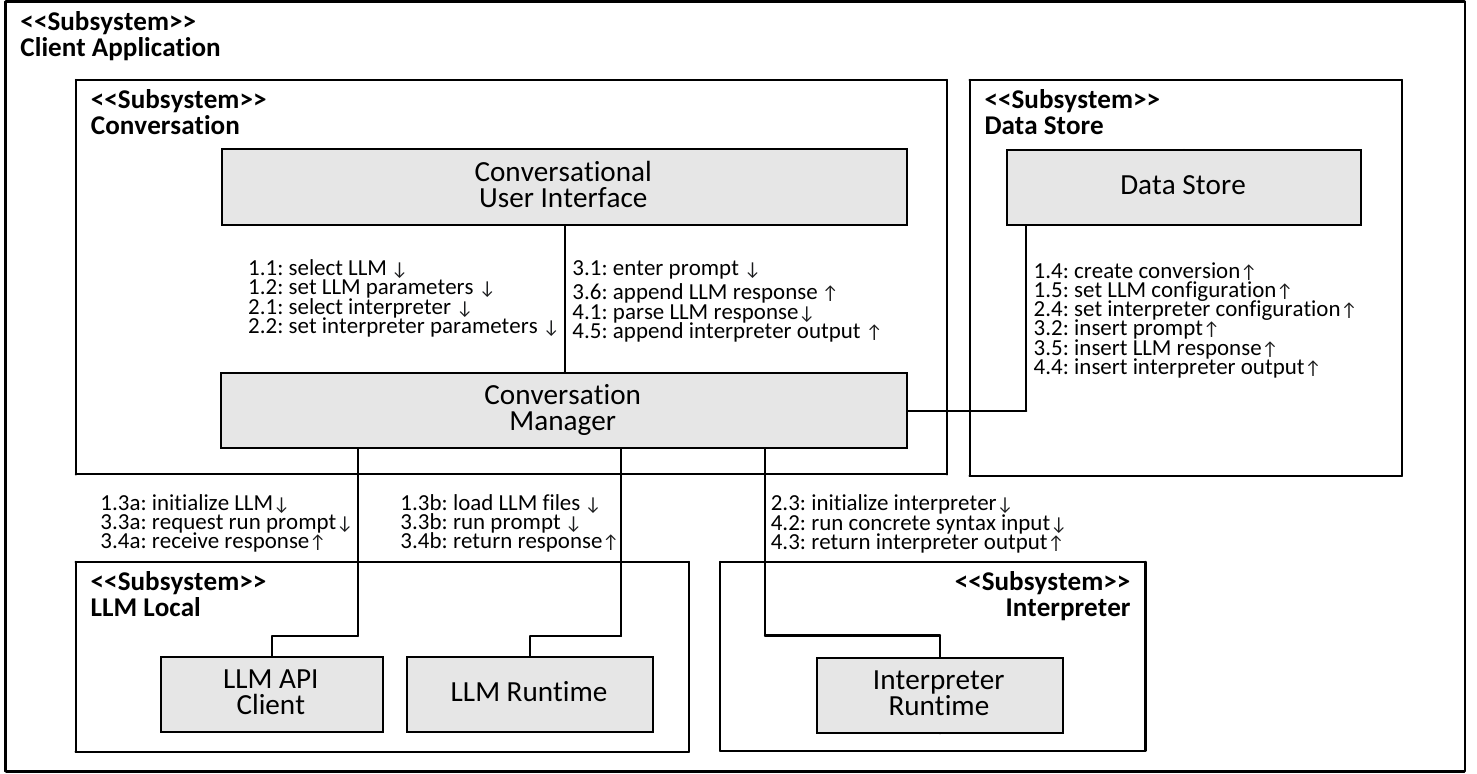}
    \caption{Architecture describing the necessary components of a client application as UML Communication Diagram. The main control flows are indicated by numbered sequences.}
    \label{fig:architecture}
    \vspace{-5mm}
\end{figure*}

Based on the concept and requirements, Figure~\ref{fig:architecture} shows components realizing a possible architecture of a client application. In the \textit{Conversation} subsystem, the main control flows are initiated through the \textit{Conversational User Interface}. According to the sequences in Figure~\ref{fig:architecture}, the user engages in the preparation of a LLM (1.1-1.5), the preparation of an interpreter (2.1-2.4), the interaction with the LLM (3.1-3.6), and the interaction with the interpreter (4.1-4.5). 

After selecting and setting up the LLM (1.1-1.2), the \textit{Conversation Manager} prepares the initiation of the conversation. Depending on the selection, either a server-side LLM is initialized through the \textit{LLM API Client} (1.3a), or client-side LLM files are loaded (1.3b) with the \textit{LLM Runtime}. Following the preparation of the LLM, the conversation and its configuration are recorded in the \textit{Data Store} (1.4-1.5). In a similar way, an interpreter is selected and set up (2.1-2.2) with the \textit{Conversation Manager}. The interpreter is initialized by the \textit{Interpreter Runtime} (2.3), assumed running locally, and with its configuration recorded in the \textit{Data Store} (2.4). 

The user starts interacting with a LLM by entering a prompt that is sent to the \textit{Conversation Manager} (3.1) and inserted in the \textit{Data Store} (3.2). Either, a server-side LLM is requested to run the prompt with the \textit{LLM API Client} (3.3a) and returns the received response (3.4a). Alternatively, a client-side LLM runs the prompt directly (3.3b) and returns the response (3.4b). The response is inserted in the \textit{Data Store} (3.5) and appended in the \textit{Conversational User Interface} (3.6).

For interacting with the interpreter, an appended response is parsed by the \textit{Conversation Manager} to detect a supported modeling language syntax (4.1). The selected interpreter is executed by running it in the \textit{Interpreter Runtime} (4.2) with the detected syntax as input. The output is returned to the \textit{Conversation Manager} (4.3), inserted in the \textit{Data Store} (4.4), and appended to the dialogue in the \textit{Conversation Manager} (4.5).

\section{Results of initial experiments}
\label{results}

This section discusses first experimental results of executing a prototype implemented for ChatGPT 4 and Llama~2 with two test cases covering the generation and interpretation of models in PlantUML and Graphviz syntax. Figure~\ref{fig:uml-gpt4-1} shows the beginning of a dialogue in the user interface of the application, implemented as a feasibility demonstration in Python 3.11\footnote{\url{https://github.com/fhaer/llm-cmi}}. 

\subsection{PlantUML test case}

At the beginning of the dialogue in Figure~\ref{fig:uml-gpt4-1}, the user-provided prompt describes the test case for generating a UML class diagram according to a scenario. Without an explicit specification of the diagram, it is the task of the LLM to create classes and relationships such as orders, order items, and customers. The aim is to observe how the case is conceptualized by GPT-4 and Llama~2.

\begin{figure*}[ht]
    \centering
    \includegraphics[width=0.96\linewidth]{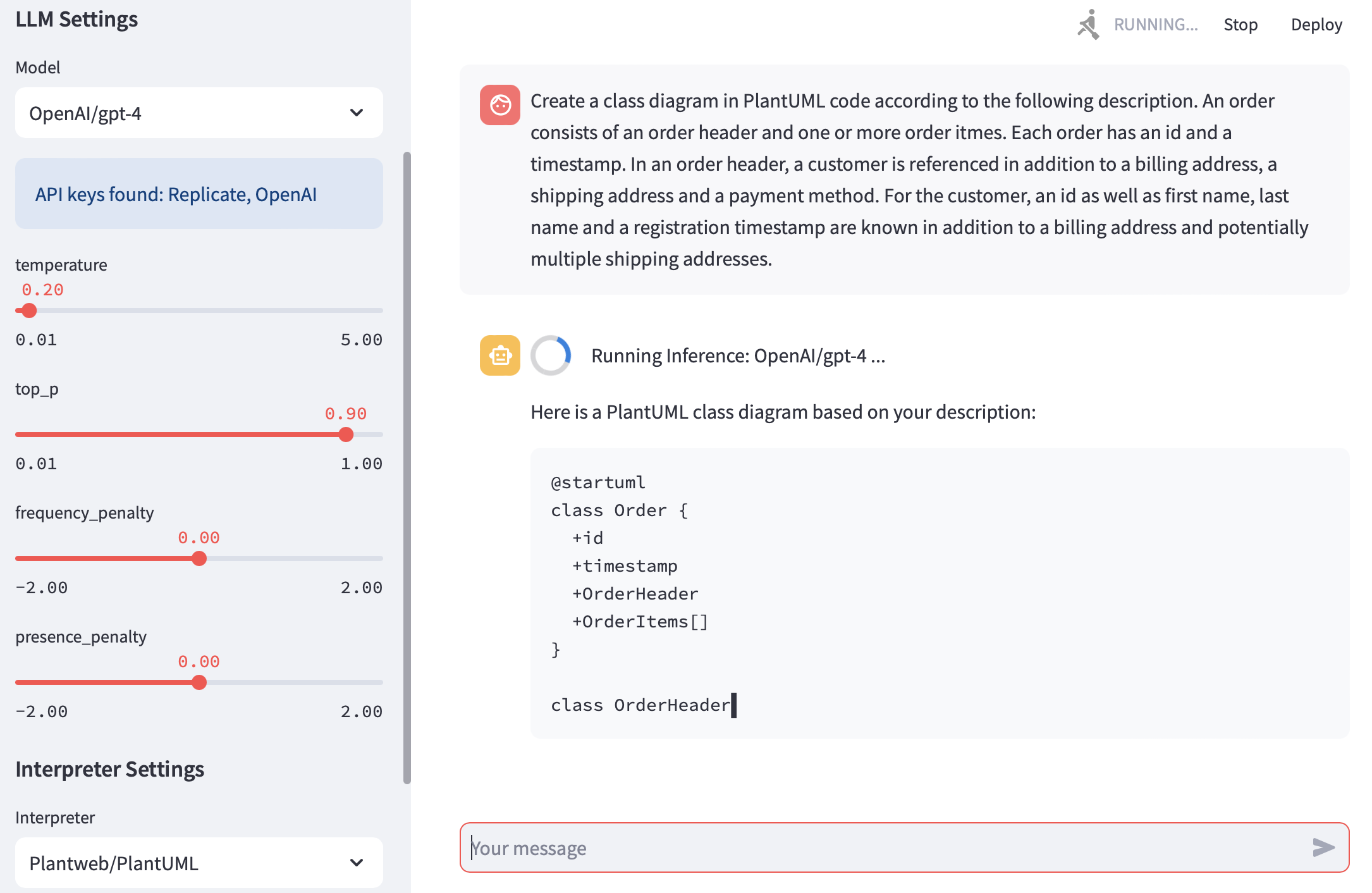}
    \caption{User interface of the prototype showing the PlantUML test case with the settings (left) and dialog (right) for GPT-4. PlantUML code is generated by GPT-4 and rendered as shown in Figure~\ref{fig:uml-gpt4-2}.}
    \label{fig:uml-gpt4-1}
    \vspace{-3mm}
\end{figure*}

At first, GPT-4 is run using the settings visible on the left-hand side in Figure~\ref{fig:uml-gpt4-1}, resulting in the diagram in Figure~\ref{fig:uml-gpt4-2}. GPT-4 created syntactically correct PlantUML code and recognized classes, attributes, and relationships with multiplicities. All classes and relationships seem appropriate for the scenario, including the classes OrderItem, PaymentMethod, BillingAddress, and ShippingAddress, for which no specific properties or attributes were described. Possibly, BillingAddress and ShippingAddress could be further generalized. Attributes use list notation, recognizing the multiplicities correctly, and specify visibility, which is always private except for OrderHeader and OrderItem[] in the Order class for no apparent reason. While the spelling is generally appropriate for software classes, the capitalization of attributes referencing classes seems slightly out of place. 

\begin{figure*}[ht]
    \centering
    \includegraphics[width=0.97\linewidth]{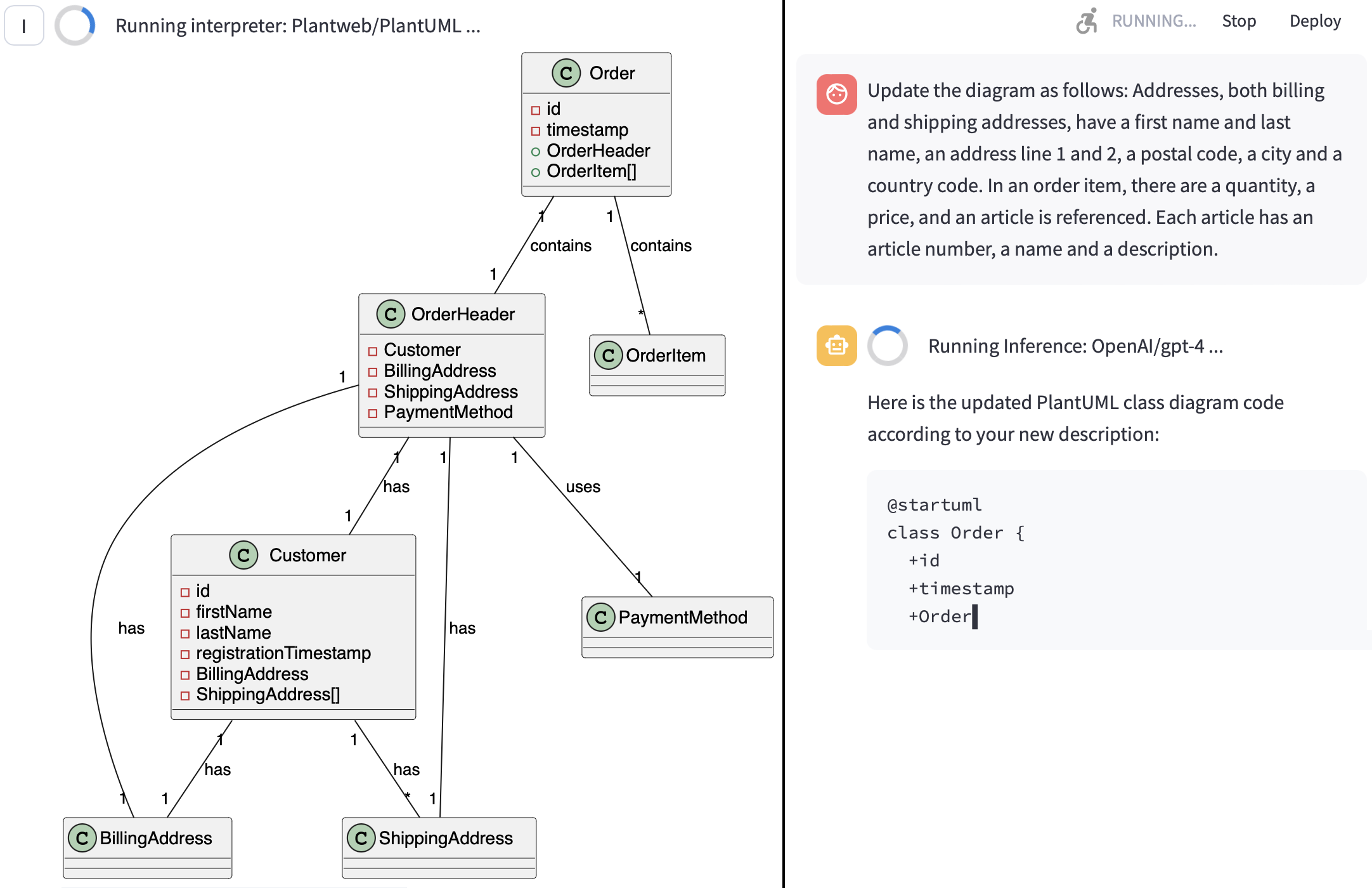}
    \caption{Continuation of the dialogue from Figure~\ref{fig:uml-gpt4-1} showing the rendered class diagram (left) and an update requested by the user. After re-generation of the syntax, the rendering in Figure~\ref{fig:uml-gpt4-3} results.}
    \label{fig:uml-gpt4-2}
    \vspace{-4mm}
\end{figure*}

Figure~\ref{fig:uml-gpt4-2} shows the continuation of the dialogue on the right-hand side, requesting to extend the diagram with articles and further descriptions of order items and addresses. After re-generating the PlantUML syntax and rendering, the class diagram shown in Figure~\ref{fig:uml-gpt4-3} is the final result of the test case. A new class Article with attributes and further attributes in existing classes were created according to the description. Furthermore, GPT-4 introduced a class Address with corresponding attributes and referenced it in the classes BillingAddress and ShippingAddress by attributes. In this way, the duplication of the address attribute was prevented without a generalization, using composition rather than inheritance. 
Variability in the responses could be observed, e.g. distinguishing between the two types of addresses in different ways and using compositions instead of associations. In a few cases, additional data types and simple operations such as getter methods were additionally created.

\begin{figure*}[ht]
    \centering
    \includegraphics[width=0.84\linewidth]{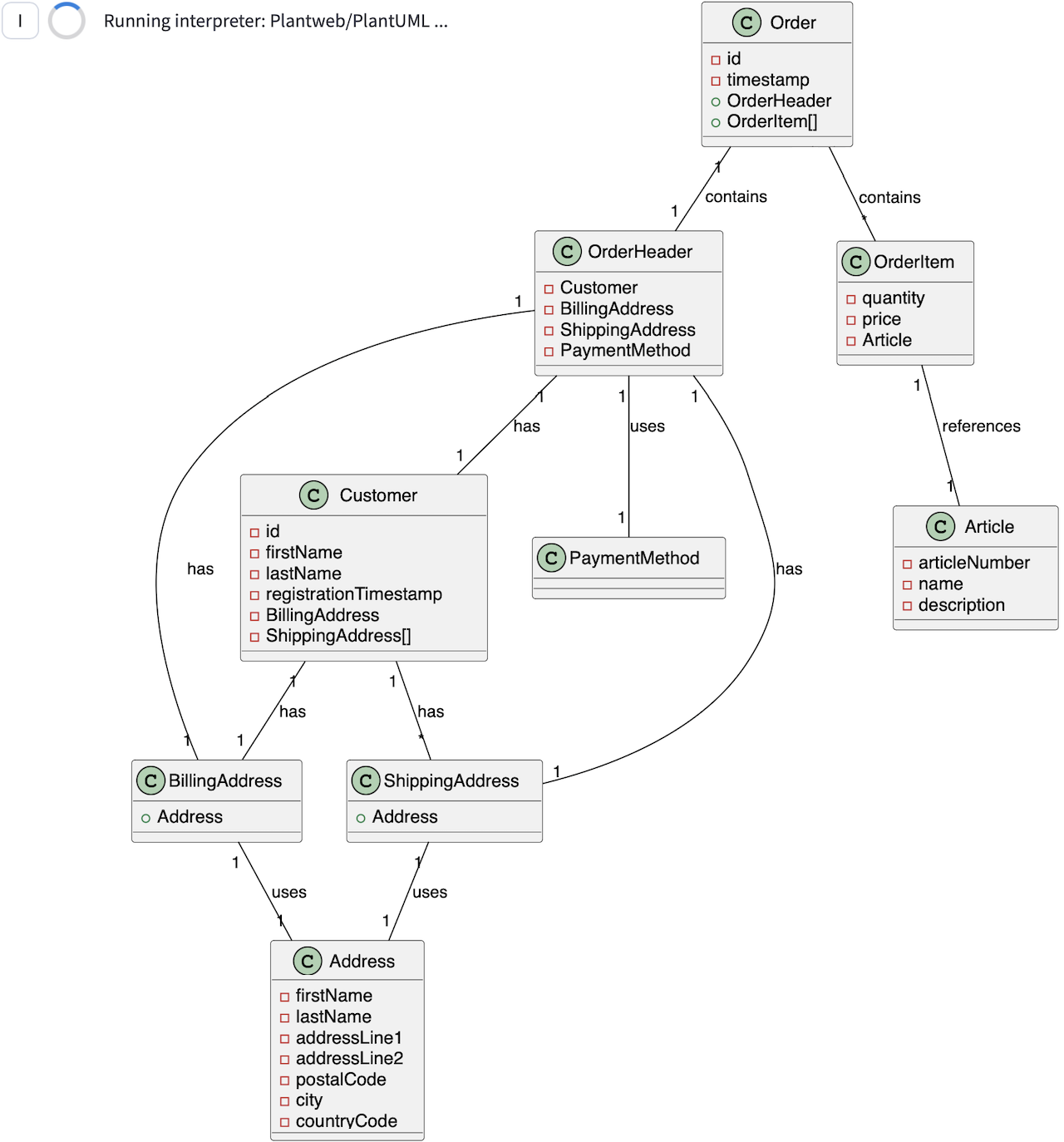}
    \caption{Continuation of the dialogue from Figure~\ref{fig:uml-gpt4-2}, showing the updated class diagram.}
    \label{fig:uml-gpt4-3}
    \vspace{-4mm}
\end{figure*}

For Llama~2, the test case is applied as before with custom settings (Appendix, Figure~\ref{fig:app-uml-llama2-1}). Also for this LLM, the result is syntactically correct PlantUML code for a class diagram, however, only classes were generated without any relationships. When asked to add relationships or associations, the LLM either did not add them or added incorrect syntax to the code. 

\begin{figure*}[ht]
    \centering
    \vspace{-2mm}
    \includegraphics[width=0.8\linewidth]{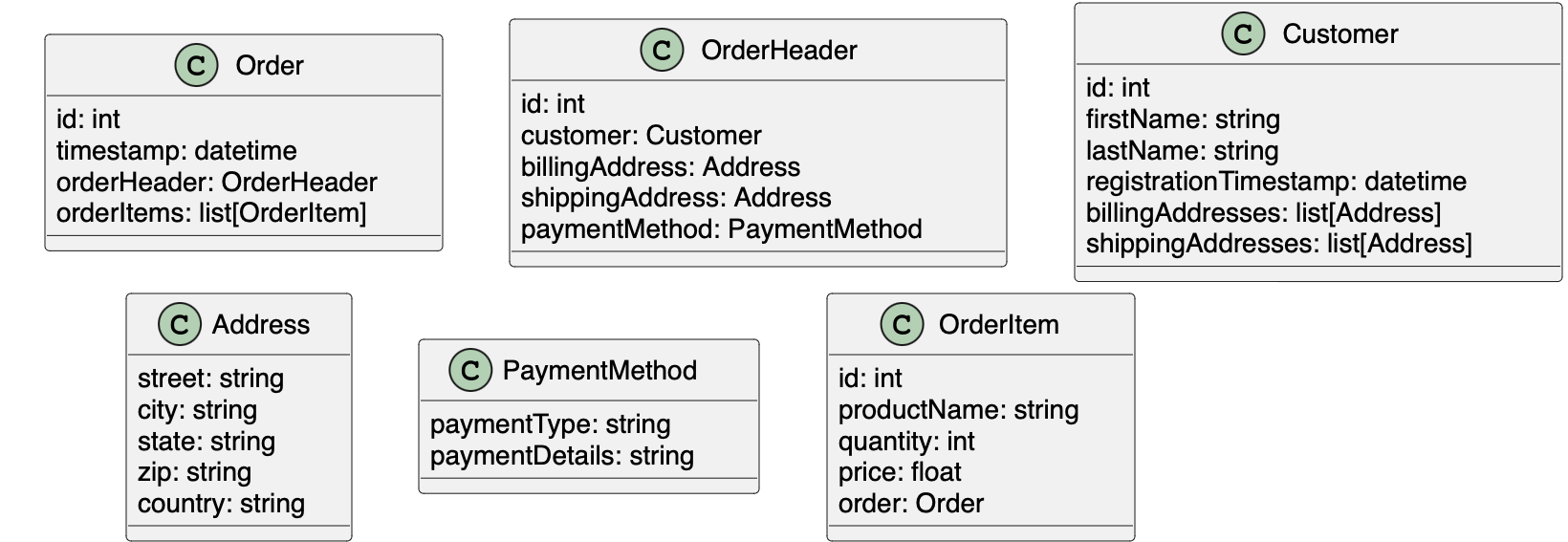}
    \caption{Rendering generated from the PlantUML test case with Llama~2 using the settings specified in the Appendix (Figure~\ref{fig:app-uml-llama2-1}). Notably, the LLM did not generate relationships.}
    \label{fig:uml-llama2-2}
    \vspace{-2mm}
\end{figure*}

Figure~\ref{fig:uml-llama2-2} shows the classes with attributes, including data types. Classes were generated similarly to GPT-4, except for the Address class which represents both billing and shipping addresses with corresponding attributes in OrderHeader and Customer. While there is no distinction on a type level and no generalization, this design also prevents the duplication of attributes. The attributes created by GPT-4 were also generated by Llama~2, representing the description correctly. However, Llama~2 also engaged in hallucination and added attributes to the OrderItem, PaymentMethod, and Address classes where the description did not indicate any specific details. Data types were correctly recognized for attributes referencing other classes and denote list types for referencing multiple classes when required by the scenario. 

The dialogue is continued as before (Figure~\ref{fig:uml-gpt4-2}), still not producing relationships (Appendix, Figure~\ref{fig:app-uml-llama2-3}). An article class was added correctly, including attributes. The scenario introduced further details for addresses and order items, requiring new attributes. These were added only, it seems, if they did not conflict with existing ones. E.g., no address line and postal code attributes were added to Address which already had street and zip attributes due to prior hallucination. Variability could be observed in slight changes to the chosen attributes and data types.

\subsection{Graphviz test case}

\begin{figure*}[hb]
    \centering
    \includegraphics[width=0.94\linewidth]{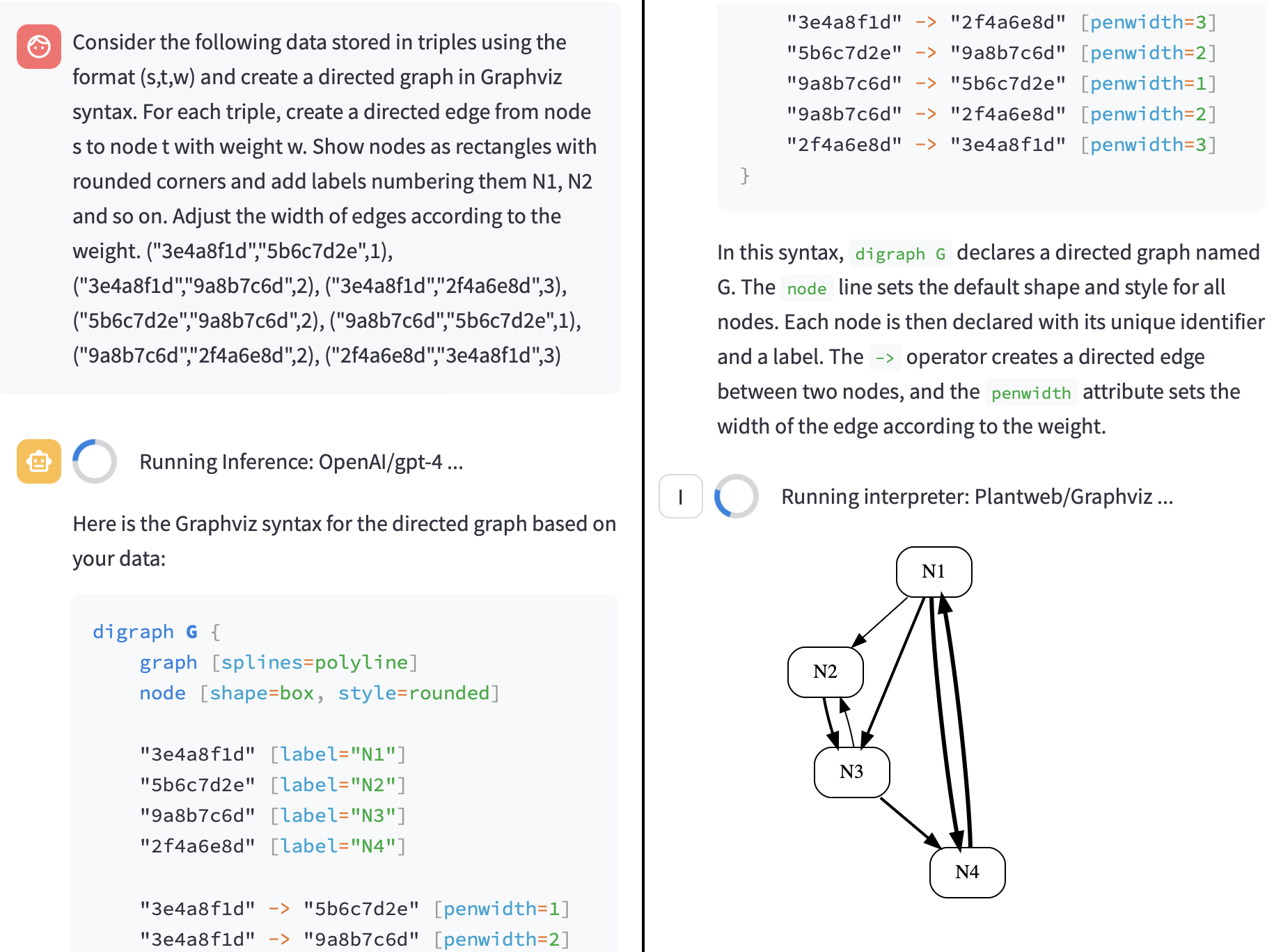}
    \caption{Graphviz test case executed with GPT-4 with the resulting rendering.}
    \label{fig:graphviz-gpt4}
    \vspace{-3mm}
\end{figure*}

In the Graphviz test case, a description for generating a directed graph from user-provided data is given to GPT-4 and Llama~2 as described in Figure~\ref{fig:graphviz-gpt4}. The aim is to observe how the LLMs construct graphs on an instance level from custom data with syntax and formatting instructions. 

\begin{figure*}[ht]
    \centering
    \includegraphics[width=1.00\linewidth]{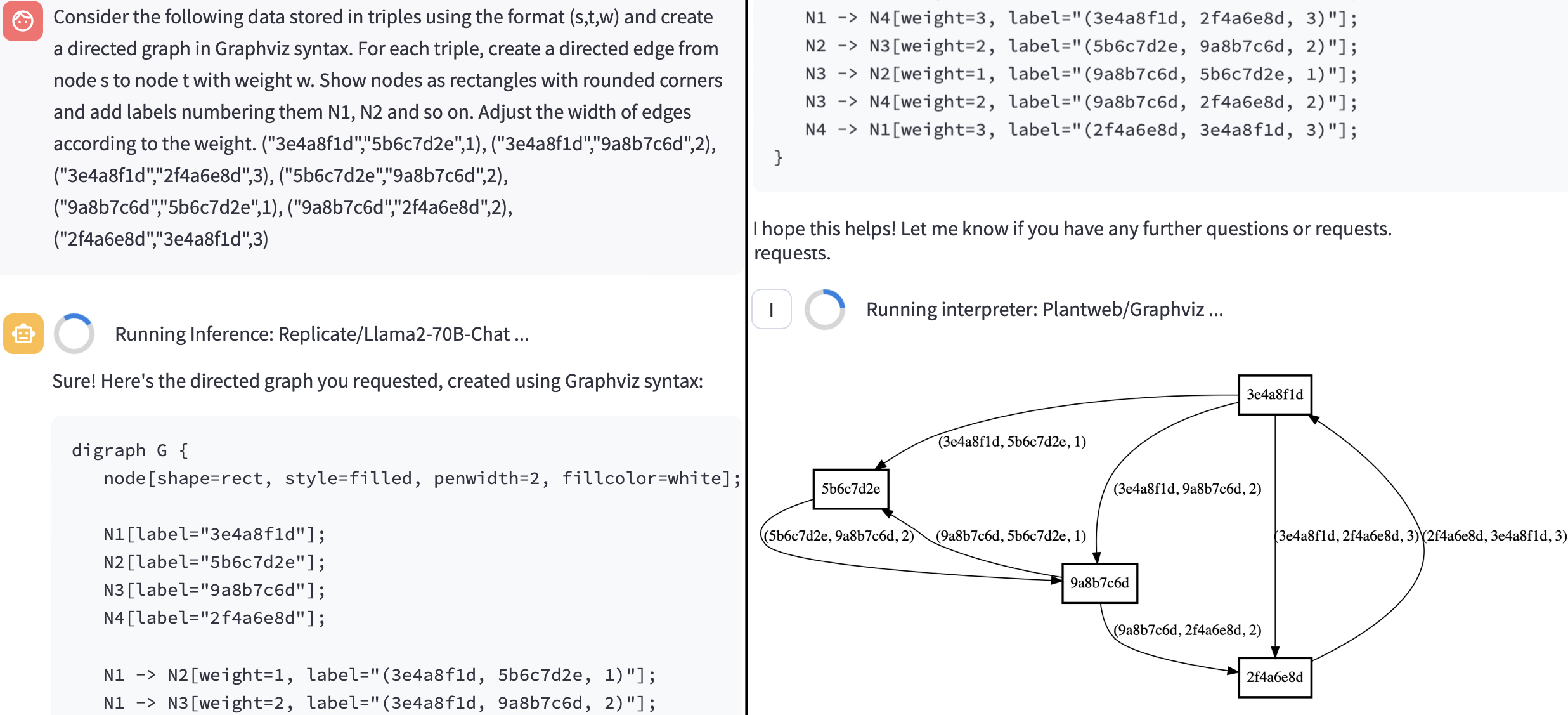}
    \caption{Graphviz test case executed by Llama~2 with the resulting rendering.}
    \label{fig:graphviz-llama2}
\end{figure*}

GPT-4 generated syntactically correct Graphviz code and created the graph as specified by the data. For the nodes, further formatting instructions for the shapes and a custom numbering format for labels were correctly implemented. As requested, edges were correctly visualized by their weight using the width; however, the weight attribute should have been specified as well. For Llama~2, Figure~\ref{fig:graphviz-llama2} shows the result with the settings applied before (Appendix, Figure~\ref{fig:app-uml-llama2-1}). The Graphviz code was also generated correctly, in terms of the syntax and the graph specified by the data. Compared to GPT-4, node names and labels are swapped since the specified numbering format was applied to the node names instead of the labels. Further, the shapes are missing the style attribute. Regarding the edges, the weight attributes were correctly set without the requested visualization of the width. Unnecessary edge labels denoting the graph data were additionally generated. Generally, the data and its syntax were recognized correctly by GPT-4 as well as Llama~2 with relatively low variability.

\subsection{Discussion}

The syntax for PlantUML and Graphviz was generally created correctly by GPT-4 and Llama~2. GPT-4 utilized the syntax to a greater extent and produced more comprehensive solutions with greater complexity, also considering details recognized from the case descriptions. For Llama~2, it remains unclear whether missing elements were unknown in their syntax or not recognized, especially in the PlantUML test case where relationships had to be inferred and were missing. In this test case, the LLM was also prone to hallucination. When given a syntax description and custom data in the Graphviz test case, both Llama~2 and GPT-4 were able to recognize it and rendered graphs on an instance level. Due to the chosen phrasing and parameters, variability was generally low, however, changes as described in the results could still be observed. As expected, parameters of the sampling influenced variability, i.e., temperature, top\_p, and top\_k. Concerning the prototype, limitations exist in its current implementation stage, not realizing the architecture in full regarding compatibility with APIs and local LLMs beyond GPT-4~and~Llama~2.

\section{Conclusion}
\label{conclusion}

In this paper, the application of a conceptual model interpreter for LLMs was explored through the creation of an architecture and prototype using GPT-4 and Llama~2 for generating and rendering models in UML and Graphviz syntax. Results encompass (1.) the components of the architecture compatible with state-of-the-art LLMs and interpreters, and (2.) initial experimental results, demonstrating the creation of models in correct syntax for GPT-4 and Llama~2 with major advantages in correctness, recognized details, and comprehensiveness for GPT-4. Especially for GPT-4, the results show that modeling iteratively in a conversational dialogue could be practical, however, further systematic evaluations need to be conducted. Future research will apply the interpreter for these evaluations with additional commercial and open source LLMs.

\begin{acknowledgments}
  This work is partially supported by the Swiss National Science Foundation project Domain-Specific Conceptual Modeling for Distributed Ledger Technologies [196889].
\end{acknowledgments}

\bibliography{main}


\appendix
\section*{Appendix}
\label{appendix}

\counterwithin*{figure}{part}
\stepcounter{part}
\renewcommand{\thefigure}{A.\arabic{figure}}


\begin{figure*}[ht]
    \centering
    \vspace{-3mm}
    \includegraphics[width=0.26\linewidth]{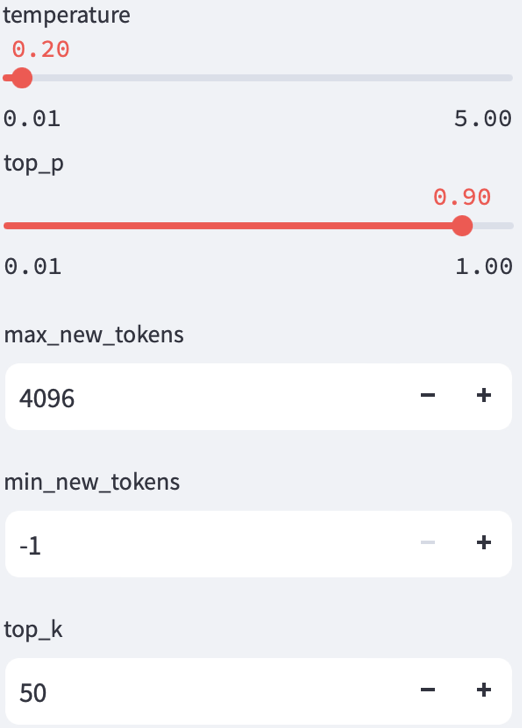}
    \caption{Settings used in the PlantUML and Graphviz test cases with Llama~2.}
    \label{fig:app-uml-llama2-1}
    \vspace{-3mm}
\end{figure*}

\begin{figure*}[ht]
    \centering
    \includegraphics[width=0.86\linewidth]{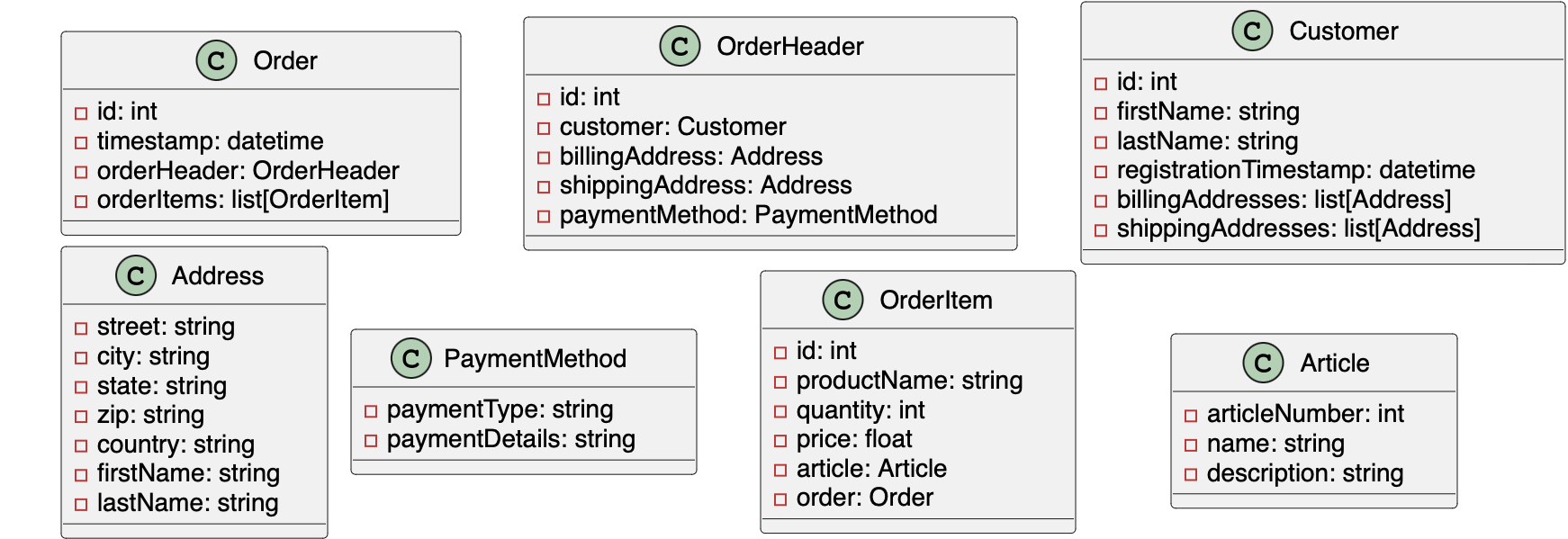}
    \caption{PlantUML rendering from Llama 2, after requesting an update of the initial diagram.}
    \label{fig:app-uml-llama2-3}
    \vspace{-3mm}
\end{figure*}

\end{document}